\documentclass[12pt,a4paper,twocolumn]{narms2}

\usepackage{graphicx}
\usepackage{subfigure} 
\usepackage{psfig}
\usepackage{timesmt}   
\usepackage{chikako}   

\makeatletter
\renewcommand{\section}{\@startsection%
{section}%
{1}%
{0mm}%
{- \baselineskip}%
{0.15\baselineskip}%
{\normalfont\normalsize}}%

\renewcommand{\subsection}{\@startsection
{subsection}%
{2}%
{0mm}%
{-\baselineskip}%
{0.15\baselineskip}%
{\normalfont\normalsize}}%
\makeatother

\begin{document}

\title{Compaction and mobility in randomly agitated granular assemblies}
\author{\large {G. Caballero}\\{\em Faculty of Science, UNAM, Mexico City, Mexico.}\vspace{20pt}\\ 
\large{J. Lanuza \&  E. Cl\'{e}ment.}\\{\em Physique et M\'{e}canique des Milieux H\'{e}t\'{e}rog\'{e}nes, Ecole Sup\'{e}rieure de Physique et de Chimie Industrielles}\\ {\em 10 rue Vauquelin, 75231 Paris Cedex 05, France.}
}

\date{}

\abstract{We study the compaction and mobility properties of a dense granular material under weak random vibration. By putting in direct contact millimetric glass beads with piezoelectric transducers we manage to inject energy to the system in a disordered manner with accelerations much smaller than gravity, resulting in a slow compaction dynamics and no convection. We characterize the mobility inside the medium by pulling through it an intruder grain at constant velocity. We present an extensive study of the relation between drag force and velocity for different vibration conditions and sizes of the intruder.}


\maketitle
\frenchspacing   


\section{INTRODUCTION}
Dynamics near jamming (glassy phase, aging, memory, intermittency) shows
amazing analogies among a variety of very different systems (colloids, dense
suspensions, foams, granular materials). Recently, several proposals have
emerged with the aim of describing in a general and unified way this common
behavior \shortcite{liu98}. With the idea of testing experimentally these,
several studies have concentrated on granular materials under vibration. It
was proposed that dense granular assemblies could be interesting models
where a new kind of fluctuation-dissipation relations is likely to extend
the classical notion of thermal temperature \shortcite{makse02}. It was also
suggested that a ''structural temperature'' could be a relevant concept that
governs the granular reorganization dynamics around a local steady state in
the jammed regime (Metha \& Edwards \citeyearNP{metha89}; Fierro et al. \citeyearNP{fierro02}). Experimentally, the dynamics of granular
assemblies have been studied under different modes of excitation either
under sinusoidal vibration or under sequences of taps. For granular
assemblies under tap, density was shown to exhibit slow compaction regime
and striking memory effects that resemble the phenomenology of spin glasses
or ferro electric glasses (Nowak et al. \citeyearNP{Nowak98}). Experimental determination of
fluctuation-dissipation relation was also proposed, but it is not clear
then, that in the weak agitation limit, the perturbing action of the probe
can be decoupled from the intrinsic dynamics of the medium (D'anna et al. \citeyearNP{danna03}).
Moreover, in a regime\ where agitation is of the order or much larger than
the acceleration of gravity, a granular assembly experiences two strongly
distinct phases: an impact phase where grains reorganize and a launch phase
where the average confining pressure is almost zero (free fall). In this
last phase the resistance to external drag is almost inexistent and thus
governs the mobility of an externally driven intruder \shortcite{zik92}. Furthermore, in all
these situations, several studies have sown that boundaries generate
convection effects that may couple strongly to the reorganization dynamics (Caballero et al. \citeyearNP{caballero04}; Philippe \& Bideau \citeyearNP{PhilippePRL}). 

In this work, we propose a new mode of energy injection (sound waves) at
high frequencies and low acceleration that produces a random agitation of
the grains and suppress the convection effects. We present experiments on
the mobility dynamics of an intruder driven in this weakly agitated medium.

\section{EXPERIMENTAL SET-UP}

\begin{figure}[hh]
\begin{center}
\includegraphics[scale=1.2]{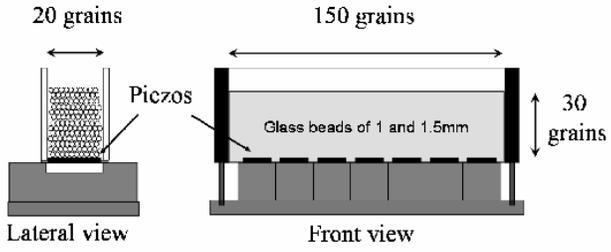}
\caption{\small Experimental setup. A glass container is closed at its bottom by seven piezoelectric transducers. The container is filled with a bidisperse mixture of glass beads of 1 and $1.5mm$ that are in direct contact with the transducers. Each piezoelectric is excited by a $400Hz$ square signal that is out of phase by $\pi$ compared to that of its neighbors.}
\label{fig:setup}
\end{center}
\end{figure}

Figure \ref{fig:setup} schematically shows the experimental setup. A
rectangular glass container is closed at its bottom by seven piezoelectric
transducers, each one glued by its extremes to a plastic base in a way that
the ceramic membrane is free to vibrate, being directly in contact with the
grains only  (figure \ref{fig:setup}, lateral view). The plastic bases of
the transducers as well as the glass container are all fixed to a main
plastic support. The container is filled with a bidisperse mixture of glass
beads of $1$ and $1.5mm$ and each piezoelectric is excited by a $400Hz$
square signal that is out of phase by $\pi $ compared to that of its
neighbors. With this setup, the agitation created in the bulk is quite
disordered and weak compared to typical experiments of vibrated granular
materials \shortcite{Nowak98,PhilippePRL,kabla} where all the container is
shaken with accelerations of the order of or greater than gravity. The
effective acceleration induced in the bulk is measured as a function of the
input voltage with an accelerometer buried in the bulk. We found an
homogeneous vibration level in the cell and a linear relation between
driving voltage and effective acceleration level  where the maximum voltage
of $10Volt$ corresponds to an rms acceleration ($\gamma _{rms}$) of $%
0.4ms^{-2}$. With such a weak excitation, the main mode excited in the
system is the rotation of grains and re-organization of granular contacts.
By tracking tracers we have verified that there is neither diffusion nor
convection, at least for times as long as 24 hours.  

\section{MOBILITY AT CONSTANT VELOCITY}

\begin{figure}[hh]
\begin{center}
\includegraphics[scale=1.2]{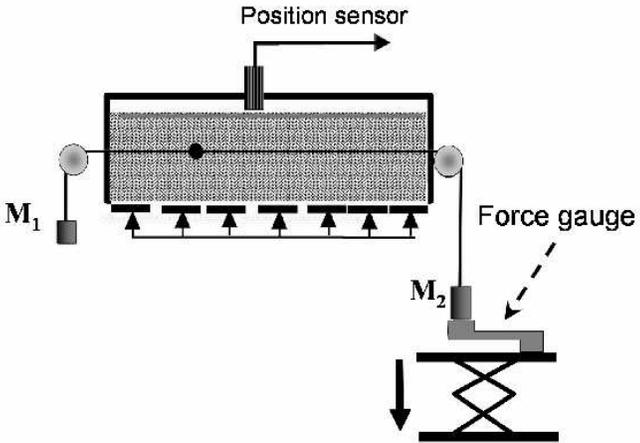}
\caption{\small An Attwood-like machine is used to measure the drag force that results of driving an intruder bead glued to a metallic thread through the granular medium.}
\label{fig:mobility}
\end{center}
\end{figure}

Figure \ref{fig:mobility} shows the experimental setup used to measure the
mobility of the system by driving an intruder grain at constant velocity. A
tense thread ($M_{1}<M_{2}$) supported by two pulleys passes through the
container filled with grains. The heaviest mass $M_{2}$ rests on a force
gauge that is attached to a step motor which moves vertically at constant
velocity. This allows to measure the drag force that resists the movement of
the intruder grain. The drag force can be measured as a function of driving
velocity, intensity of vibration and size of the intruder.

\begin{figure}[hh]
\begin{center}
\includegraphics[scale=1.2]{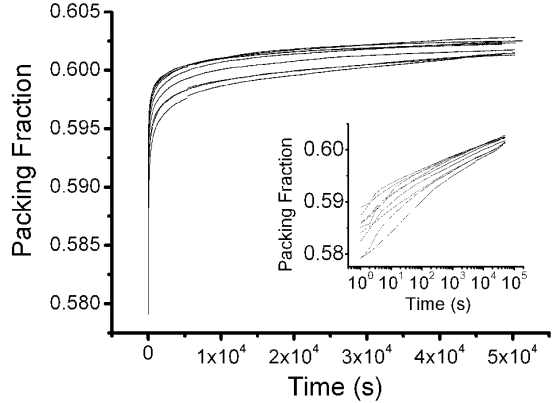}
\caption{\small Evolution with time of the packing fraction of the granular pile vibrated at the maximum intensity ($\gamma_{rms}=0.4m/s^2$). There are shown ten independent experiments. Inset: Same data but with the time in logarithmic scale.}
\label{fig:packing}
\end{center}
\end{figure}
The packing fraction of the piling is obtained by placing a metallic lid on
the surface of the pile and two inductive position sensors that measured the
lid-sensor separation with a resolution of the order of micrometers (figure 
\ref{fig:mobility}). Figure \ref{fig:packing} shows the packing fraction
evolution with time for ten different experiments vibrated at the maximum
intensity. After fourteen hours of vibration a stationary state is not
reached. However, since the compaction is logarithmic, at that stage the
packing fraction can be considered as almost constant during the mobility
experiments. Noteworthy is the reproducibility of the compaction curves for
independent experiments, which allows us to say that mobility measures
presented below were made at equal conditions.

The mobility experiment protocol was the following : $265$ $g$ of glass
beads of $1$ and $1.5mm$ were poured in the container trying to make the
upper surface as horizontal as possible without tapping the container. Then,
the thread with the intruder grain was made to pass through the grains $16mm$
deep from the surface of the pile (the total height was around $40mm$). The
metallic lid was gently placed on the surface of the pile and the
piezoelectric transducers were turned on at their maximum intensity ($\gamma
_{rms}=0.4m/s^{2}$). After 14 hours of vibration the intruder grain was
driven at constant velocity along ten centimeters.

\section{RESULTS}

\begin{figure}[hh]
\begin{center}
\includegraphics[scale=1.2]{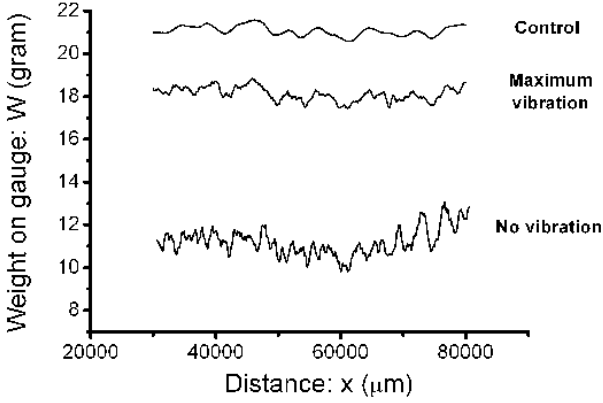}
\caption{\small Weight measured by the gauge while driving a $9mm$ bead at $10\mu m/s$ at strongest vibration and without vibration. It is also shown a control experiment where the bead was driven through the same length but without grains in the container.}
\label{fig:WvsD}
\end{center}
\end{figure}
For these experiments an intruder bead was glued to a $100\mu m$ metallic
thread and we use two bead sizes of diameters $d_{1}=9$ and $d_{2}=2mm$. We
also monitor the drag of the metallic thread alone.  In figure \ref{fig:WvsD}
we show two examples of drag force measurements as a function of distance $x$
at a constant driving velocity of $10\mu m/s$. Measurements were performed
without vibration and under a $0.4m/s^{2}$ acceleration. We also display a
control experiment where bead and thread are driven at $100\mu m/s$ through
the same path but without grains in the container. It represents the
reference line and the intrinsic noise level of the Attwood setup. The mean
drag force $<F_{d}>_{x}$ is obtained from $<F_{d}>_{x}=g(<W_{0}(x)-W(x)>_{x})
$, where $g$ is the acceleration of gravity, $W(x)$ is the weight measured
by the gauge at position $x$ while driving the intruder bead through the
grains, $W_{0}(x)$ is the weight measured at $x$ in the control
experiment and the average is taken for displacement $2<x<8mm$. We already
observe that the injection of sound waves  strongly modifies the penetration
resistance in the granular assembly.

\begin{figure}[hh]
\begin{center}
\includegraphics[scale=1.2]{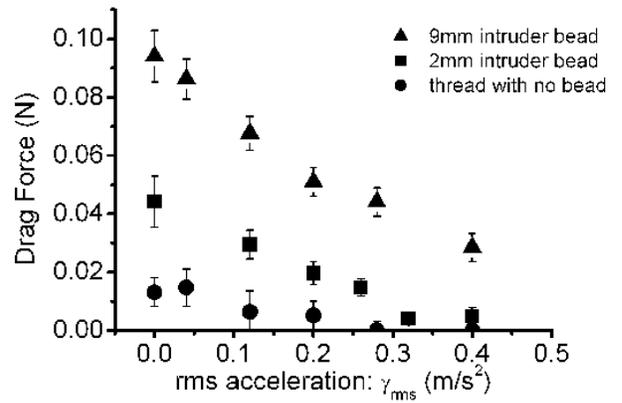}
\caption{\small Drag force at $10\mu m/s$ as a function of vibration intensity which is expressed in terms of the rms acceleration of the grains in the pile.}
\label{fig:FvsAcc}
\end{center}
\end{figure}

Figure \ref{fig:FvsAcc} shows the drag force as a function of the vibration
intensity or, equivalently, the rms acceleration of the grains in the bulk.
Driving velocity was of $10\mu m/s$ and measurements were made for the
thread alone  and for the two bead sizes: $d_{1}=9mm$ and $d_{2}=2mm$. For
the thread alone and the $2mm$ intruder bead the force drops almost linearly
to zero. This threshold was not reached for the $9mm$ bead since $0.4ms^{-2}$
is so far the upper limit of rms acceleration. We might expect beyond this
limit where drag force almost vanishes, a new  regime as far as rheology is
concerned and a more viscous -like regime. This issue  under current
investigation. But in this report, we concentrate on the regime observed at
lower accelerations and for the larger intruder bead.  For this systemn we
obtained systematic measurements of the drag force as a function of the
driving velocity.

\begin{figure}[hh]
\begin{center}
\includegraphics[scale=1.2]{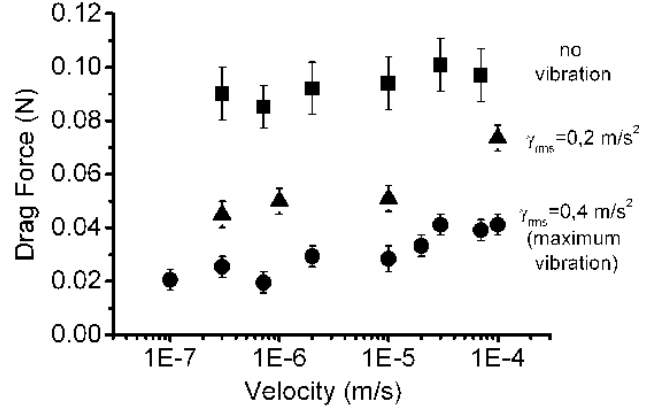}
\caption{\small Drag force vs. driving velocity for a $9mm$ intruder bead at different vibration intensities.}
\label{fig:FvsVel9mm}
\end{center}
\end{figure}
Figure \ref{fig:FvsVel9mm} shows drag force measurements  for the $9mm$ bead
with and without vibration. For the maximum vibration ($\gamma
_{rms}=0.4m/s^{2}$) the drag force increases logarithmically with velocity
on three decades: $F\simeq F_{0}\left( 1+\beta \ln \frac{V}{V_{0}}\right)$
, with $F_{0}=2.10^{-2}$,$V_{0}=10^{-7}$ and $\beta =0.144$. For $\gamma
_{rms}=0.2m/s^{2}$ the relation seems to be also logarithmic, though there
are not many points. It is difficult to say something for the no vibration
case because the variation in the value of the drag force, if any, is small
compared to the error bar. In similar experiments in non vibrated granular
materials, Albert et al. \shortcite{schifferPRL99} find a drag force
independent of driving velocity, though they work with higher velocities
than ours. In contrast, for small velocities but in a two dimensional setup, \shortcite{behringer} find a logarithmic increment of the drag force
with velocity. The grains used in this latter experiment were made of a
photoelastic material. J. Geng and R. Behringer propose that there is a
relaxation time associated to the smoothness of the grains which explains
the rate dependence of the drag force. In our experiment as well as in that
of Albert et al., the rigidity of glass beads implies that there is not such
a relaxation time and the drag force becomes independent of drag velocity.
Thus, the logarithmic relation for vibrated experiments in figure \ref
{fig:FvsVel9mm} suggest that vibration introduces a relaxation time to the
system and, consequently, a dependence of the force on velocity.

\section{CONCLUSIONS}

The relation between drag force and velocity of an object moving through a
granular system has been studied for velocities and vibration conditions
different from any previous work \shortcite{zik92,schifferPRL99,behringer}.
Vibration by an array of piezoelectric transducers allowed a weak and
disordered agitation where the main mode excited is the rotation of the
grains and tiny local reorganizations of a grain environment (contact
distribution). The packing fraction exhibits a slow logarithmic growth when
the pile is vibrated at the maximum available intensity for fourteen hours without reaching a stationary state. We have not yet performed compaction experiments for longer times so we do not even know the order of magnitude of the time required for stationarity or the possible effect that this could have on our rheology measures. So far, the only thing that we know is that all mobility experiments are done under equivalent density conditions. 

In spite of accelerations much lower than the acceleration of gravity it was
shown that the drag force depends significantly on the vibration intensity.
Here we evidence a regime were at a constant energy input, the drag force
increases in a logarithmic way with velocity. This result suggests an
activated dynamics in analogy with solid on solid friction models 
\shortcite{rice,heslot} and/or with glassy transition on thermal systems 
\shortcite{ediger96}. In the friction scenario the relevant parameter is an
effective friction coefficient while in the glass analogy it is the
viscosity.  Both parameters are related to the temperature of the system,
which means that the correct determination of an effective viscosity or a
friction coefficient in granular systems would allow the definition of an
effective granular temperature. In the regime we probe, we do not evidence a
linear force/velocity relation that could validate a fluctuation dissipation
approach to describe the local dynamics.

\vspace{5pt}

{\bf Acknowledgments} This project is part of ECOS M03P01 and GC is supported 
by CONACYT and DGEP.

\newpage

\bibliographystyle{chikako}      
\bibliography{GCaballero} 

\end{document}